\documentclass[prb,aps,preprint, showpacs]{revtex4-1}   

\usepackage{graphicx}

\begin{document}
\bibliographystyle{prsty}

\title{Emergence of current branches in a series array of negative differential resistance circuit elements}
\author{Huidong Xu and Stephen W. Teitsworth}

\affiliation{Duke University, Department of Physics, Box 90305,
Durham, NC 27708-0305, USA}

\date{ \today  }

\begin{abstract}
We study a series array of nonlinear electrical circuit elements
that possess negative differential resistance and find that
\emph{heterogeneity} in the element properties leads to the presence of multiple branches in
current-voltage curves and a non-uniform distribution of voltages
across the elements. An inhomogeneity parameter $r_{max}$ is introduced to
characterize the extent to which the individual element voltages deviate from one another, and it is found to be strongly dependent on the rate of change of applied voltage. Analytical
expressions are derived for the dependence of $r_{max}$ on voltage ramping rate in the limit of fast ramping and are
confirmed by direct numerical simulation.

\end{abstract}
\maketitle

Negative differential resistance (NDR) - in which an increasing
applied voltage causes a \emph{reduced} electrical current flow -
occurs in a number of electronic transport systems, for example,
tunnel diodes,\cite{Esaki58} semiconductor double barrier quantum
well structures,\cite{Goldman87} and molecular bridges.
\cite{Reed99, RLiu06} When several such elements are placed in a
series array, forming a complex system, the NDR of individual
elements often leads to striking self-organization effects
associated with a spatially non-uniform distribution of electric
field, despite the fact that the elements are almost identical.  An
important example of this behavior is provided by semiconductor
superlattices consisting of a series array of quantum wells (e.g.,
composed of GaAs) separated by potential barriers (e.g., composed of
AlAs). In the case of \emph{doped} superlattices with relatively
wide barrier layers (i.e., weakly-coupled superlattices), one may observe multiple current branches in
current-voltage ($I-V$) characteristics with the number of branches
approximately equal to the number of periods of the superlattice.
\cite{Choi88, Grahn91} These branches are associated with a static,
non-uniform electric field configuration in which some periods of
the structure are in a low-field domain and the others in a
high-field domain.\cite{SLReview, Schoell2001} Current branching
and/or non-uniform electric field distributions have been reported
for other spatially-periodic systems such as quantum cascade laser
structures, \cite{LuS06} multiple-quantum well infrared detectors,
\cite{SchneiderS06} Si nanocrystal structures, \cite{Chen07} and
two-dimensional transport in laterally patterned GaAs quantum
wells\cite{Pan08}. It is remarkable that, despite the microscopic
differences between these systems, the observed behavior
is similar. This naturally raises two questions: What are the
minimal ingredients needed to observe current branch formation?  What is the
role of heterogeneity in the individual element properties on this
behavior?

In this paper, we address these questions by introducing a nonlinear
circuit model consisting of a voltage-biased series array of $N$
ideal negative differential resistance elements, each connected in
parallel to a capacitance as shown in Fig.~1a. We find that
\emph{heterogeneity} in the element properties leads to the formation of multiple
branches in current-voltage curves and a non-uniform distribution of
voltages across individual elements.

The model is defined as follows.  Each element of the series array is assumed to have an intrinsic $I-V$ curve of the form
$I_i(v_i)$ with a typical example shown in Fig.~1b. \cite{footnote1}
The total electrical current is
\begin{equation}
I=I_i(v_i)+C_i \dot{v}_i, \label{Ji}
\end{equation}
for $i = 1,2,...N$, where $v_i$ denotes the voltage across the
$i$-th element and $C_i$ is the parallel capacitance associated with
the $i$-th element. The total applied voltage to the array is
$V=\sum_{j=1}^{N}v_j$. Dividing both sides of Eq. (\ref{Ji}) by
$C_i$ and summing over $i$ allows to express the total current as
\begin{equation}
I=C_{tot}\dot{V}+C_{tot}\sum_{i=1}^{N}I_i/C_i, \label{sumi}
\end{equation}
where $C_{tot}^{-1}\equiv\sum_{i=1}^{N}C_i^{-1}$ denotes the inverse total
equivalent series capacitance.  Relabeling the summation in Eq.
(\ref{sumi}) as over $j$ and substituting the result back into Eq.
(\ref{Ji}), the circuit model equations can be written in the
following form:
\begin{equation}
 \dot{v_i}=\frac{C_{tot}}{C_i}\dot{V}+\sum_{j=1}^{N} K_{ij}
 (I_j(v_j)-I_i(v_i)),
\label{TDmodel}
\end{equation}
where $K_{ij}=\frac{C_{tot}}{C_i C_j}$. Equation (\ref{TDmodel}) has
the form of an $N$-dimensional dynamical system subject to a
constraint on the total voltage. The constraint is built into the
structure of the model, as seen by summing Eq. (\ref{TDmodel}) over
all $i$ and using the symmetry property $K_{ij}=K_{ji}$. The
structure of Eq. (\ref{TDmodel}) is equivalent to a limiting case of
a standard rate equation model for superlattices, when the doping
level in each quantum well is very large. \cite{SLReview, XuT07, Xu2010}

In this paper, we focus on the case of \emph{linear ramping} (with
ramp time $T$) in the total applied voltage $V$, so that $V$
increases linearly between values $0$ and $NV_{max}$ and
$\dot{V}=N\frac{V_{max}}{T}\equiv N\alpha$.  The first term in Eq.
(\ref{TDmodel}) describes the effect of changing total applied
voltage, while the second term describes a global coupling between
individual circuit elements.  When the ramping rate is large, the effect of the coupling term is expected to be small.  On the other hand, when the ramping rate is small, the coupling term plays a decisive role in the observed behavior.  The coupling term displays either
positive or negative feedback depending on the state of each
element. Considering pairs of elements, they attract one another
when both are on a stable branch of the $I-V$ curve (i.e., Regions I
and III in Fig. 1b), and repel if they are both on the unstable part
(Region II). If one element is in a stable region and the other is
in the unstable region, the elements may either attract or repel
depending on their relative current values.

If the coupling term in Eq. (\ref{TDmodel}) is neglected, one finds
an \emph{uncoupled} state in which the element voltages increase
independently. Starting from the initial condition $v_i^{(0)}(0)=0$
for all $i$, typical for experimental measurements, \cite{SLReview, LuS06,
Pan08} the voltage of each element is given by $v_i^{(0)}=V\times
{C_{tot}}/{C_i}$. Since the $C_i$ are assumed to have a small dispersion, the
uncoupled state is associated with a field profile that is nearly
uniform. (Here, we use the term field to refer to the spatial distribution of element voltages.)
When the coupling term is nonzero, the system deviates from
the uncoupled state and the field profile becomes non-uniform. This
behavior is usefully characterized by introducing a \emph{field
inhomogeneity parameter} $r_{max}$ defined by:
\begin{equation}
r_{max}=\max \{r(t): 0<t<T\}, \label{r}
\end{equation}
where
\begin{equation}
r(t)=\frac{1}{N}\sum_{j=1}^{N}\left(\frac{v_j-v_j^{(0)}}
{v_j^{(0)}}\right)^2.
\label{rt}
\end{equation}
The quantity $r(t)$ expresses the time-dependent level of field
nonuniformity in the system for ramp time $T$, while $r_{max}$ gives
the maximal degree of nonuniformity during the entire ramping
process and associates a single value to the entire process. For the
uncoupled state, $r_{max}$ tends to zero; however, when the effect
of the coupling term is large, $r_{max}$ assumes a value of order 1.

If the total voltage is held constant, the system always relaxes to
a state in which the currents in the different elements are identical. This corresponds to a stable
fixed point of the system with $\dot{V}=0$.  If the average applied
voltage, $V/N$, falls in the NDR region of the single element $I-V$
curve, there are \emph{multiple} fixed points corresponding to
distinct arrangements of individual element voltages.  For a
completely homogeneous system, such that all elements are
\emph{identical}, i.e., $C_i=C$ and $I_i(v)=I_0(v)$ for all $i$,
these fixed points are degenerate corresponding to the same overall
device current. \cite{Neu08}  When the homogeneous system is
subjected to a ramped voltage, starting from the initial state
$v_i(t=0)=0$ for all $i$, the element voltages remain identical
throughout the ramp process, i.e., $v_{i}(t)=V(t)/N$, and the
overall $I-V$ curve has a similar shape to that of a single element
\cite{IC}.

When heterogeneity is introduced into the system, current branches
emerge in the limit of slow ramping as shown in the $I-V$ curve of Fig.~2a.
The variance in the individual element $I-V$ characteristics is expressed as
$I_i(v_i)=I_0(v_i)(1+\epsilon_i)$, where the $\epsilon_i$'s are
independent and identically distributed (i.i.d.) random numbers,
distributed normally with mean zero and standard deviation
$\sigma_I$, i.e., $\sigma_I^2 \equiv \frac{1}{N}\sum_{i=1}^N
\epsilon_i^2$. \cite{footnote2}  We have also explored the effect of
capacitance variation and variation due to electrical noise in
individual elements, and find that the qualitative behavior is the
same as that observed when the only variance is in the element $I-V$
curves. \cite{XuT08}

When the ramp time $T$ is \emph{large}, the system exhibits
well-defined current branches in the static $I-V$ curves, shown in
Fig. 2a for $N = 8$.  As the total voltage ramps higher, the
elements pass very rapidly from Region I to III one by one, and the current exhibits an abrupt step down for each such passage.\cite{smallN} The
parameter $r_{max}$ takes a value of order 1, indicative of a
non-uniform distribution of element voltages. Figure 2b shows the
corresponding contour plot in which the current level is plotted in
gray scale vs. element number $i$ and total voltage $V$. The clear horizontal
bands indicate that, for all parts of the ramping process, the
individual device currents are identical to one another.

As the ramp time decreases, the element current levels become somewhat different from one another, see Fig.~2c. In this case, more than one element can jump
to Region III at the same time. The field distribution deviates less
from the uniform state, the current branches are rounded and smaller
in number and amplitude, and the abrupt jumps between current
branches disappear. A similar rounding of experimental current
branches versus ramping rate has been reported in weakly-coupled
GaAs/AlAs semiconductor superlattices.\cite{RogoziaT02} The
$r_{max}$ value also decreases from its large $T$ value. Figure 2d
shows the corresponding current contour plot in which the horizontal bands are still evident, but
interrupted by localized dark areas that correspond to the passage
of individual or pairs of elements through Region II.

For fast ramping, the element currents are essentially independent of one another, and the element voltages pass through
the NDR region simultaneously (Fig. 2e). The $r_{max}$ parameter
approaches zero, implying that the system behavior is very close to
the uncoupled state described previously. The $I-V$ curve of the full array follows
closely that of an individual element, $I_0(v)$. The contour plot,
Fig. 2f, shows smooth behavior and horizontal features are absent.
We have also considered ramping from different initial states
\cite{XuT08} as well as more elaborate circuit array models - e.g.,
including small series inductance and parallel capacitance with each
nonlinear element \cite{Brown09} - and find a qualitatively similar 
behavior as described above.

Plotting the value of $r_{max}$ versus ramp time $T$ for several
different $N$ values in Fig.~3a indicates that the transition from
uncoupled to fully coupled behavior is a smooth transition with an onset
that is \emph{independent} of system size $N$. For large ramp time,
$r_{max}$ approaches an asymptotic value that is $N$-dependent. As
$N \to \infty$, $r_{max}\to 2.85$, a value that depends only on the shape of the
single element function $I_0(v)$. \cite{XuT08} For the parameters
used here (i.e., $C=200$ nF and $\sigma_I=0.1$), the transition from uncoupled to coupled behavior occurs
as the ramp time increases from $\sim$ 100 $\mu$s to $\sim$ 10 ms, a
timescale range that is significantly greater than the zero-bias
characteristic $RC$ time constant associated with an individual
element, i.e., $\tau = C dI_0/dv\mid_{v=0}\simeq 2$ $\mu$s.

To better understand this behavior, we investigate the effect of
perturbations about the uncoupled state of the form
\[
v_i=v_i^{(0)}+\delta v_i \equiv v+\delta v_i, \label{v}
\]
where $v_i^{(0)}(t)=\alpha t$ denotes the uncoupled state solution.
Substituting this form into the dynamical model, Eq.
(\ref{TDmodel}), one finds the following system of differential equations valid to first order in the $\epsilon_i$'s,
\begin{equation}
\dot{\delta v_i}\simeq\frac{1}{N C}\{ I_0(\alpha t)\sum_{j=1}^N (\epsilon_j-\epsilon_i)+I'_0(\alpha t)\sum_{j=1}^N (\delta v_j-\delta v_i)\}=-\frac{I_0(\alpha t)}{C}\epsilon_i-\frac{I_0'(\alpha t)}{c}\delta v_i,
\label{diffeq}
\end{equation}
where $I_0'\equiv dI_0/dv$.  Equation (\ref{diffeq}) can be solved explicitly to yield the following expression for $\delta v_i$:
\begin{equation}
\delta v_i  =  -\frac{\epsilon_i}{C \alpha}\,
\mathrm{e}^{-\frac{I_0(v)}{C
\alpha}}\int_0^{v}I_0(v')\,\mathrm{e}^{\frac{I_0(v')}{C\,\alpha}}dv',
\label{dvcal}
\end{equation}
with the initial conditions $\delta v_i(0)=0$ for all $i$.

For large values of ramping rate $\alpha$, the two exponentials in Eq.
(\ref{dvcal}) are approximated by unity, and it immediately follows
that 
\[
\delta v_i(t)=-\frac{\epsilon_i}{c}\int_0^t I_0(\alpha t')dt'=-\frac{\epsilon_i}{c\,\alpha}\int_0^{v} I_0(v')dv' \equiv -\frac{\epsilon_i}{c\, \alpha} P(v).
\]
Substituting this result into the definition of $r_{max}$, cf. Eqs. (\ref{r}) and (\ref{rt}), allows us to write
\[
r_{max}=\frac{\sigma_I^2 T^2}{c^2
V_{max}^2} \mathrm{max}(P(v)/v)^2
\]
which demonstrates that $r_{max}$ is proportional to $T^2$ with a coefficient of proportionality
that is independent of $N$; this is confirmed in Fig. 3a. 

For somewhat
smaller $\alpha$ values, and provided the local maximum of the
element $I-V$ curve is sufficiently sharp, Eq. (\ref{dvcal}) can be
evaluated using a saddle point method to write $\delta v_i \approx
\frac{\epsilon_i I_M
\sqrt{\pi}}{\sqrt{\beta\,c\,\alpha}}\mathrm{e}^{\frac{I_M-I_0(v)}{C
\alpha}}$, where $\beta=-\frac{I''(v)}{2}\mid_{v_M}>0$. Inserting
this result back into the definition of $r(t)$, cf. Eq. (\ref{rt}),
gives

\[
r(v)=\frac{\pi \sigma_I^2 I_M^2 T}{\beta C V_{max}
v^2}\mathrm{e}^{\frac{2(I_M-I_0(v))}{C \alpha}}.
\]
In the preceding equation, we note that the
exponential term takes a maximum value if we set $v=v_m$, so that
$I_M-I_0(v)=I_M-I_m\equiv \Delta I$, where $v_m$ and $I_m$ are the
voltage and current coordinates of the local \emph{minimum} in the
element $I-V$ characteristic, cf. Fig. 1b.  This allows us to write the maximum value of $r$ over the entire ramp process as
\begin{equation}
r_{max}=\frac{\pi \sigma_I^2 I_M^2 T}{\beta C V_{max}
v_m^2}\mathrm{e}^{\frac{2\Delta I}{C \alpha}}. \label{rmax}
\end{equation}

Figure~3b plots the values of $r_{max}$ calculated from both Eqs.
(\ref{dvcal}) and (\ref{rmax}) and compares them with the numerical
results for $N = 400$, demonstrating that there is a range of ramp
times (around $10^{-4}$ seconds) for which the asymptotic
expression, Eq. (\ref{rmax}) follows closely the transition from uncoupled to coupled behavior.
In order to develop an analytic criterion for
the onset of coupled behavior - and associated field non-uniformity, we define a characteristic value of
ramp time $T_{th}$ that sits in this range. Denoting the corresponding
characteristic value of $r_{max}$ by $r_{th}$, we have $\ln
r_{th}=2\ln \sigma_I + \ln \left( \frac{\pi I_M^2}{C \beta
v_m^2}\right)-\ln \alpha+\frac{2 \Delta I}{\alpha C}$. Solving for
$T_{th}$, we can write
\begin{equation}
T_{th}=-\frac{C\,V_{max}}{\Delta I}\mathrm{ln}\sigma_I + K,
\label{Asymp}
\end{equation}
where $K = \frac{C V_{max}}{\Delta I}\left(\frac{1}{2}\ln
\frac{r_{th} \beta C V_{max}v_m^2} {\pi
I_M^2}-\frac{1}{2}\ln{T_{th}}\right)$ is slowly varying. Equation
(\ref{Asymp}) shows explicitly the dependences of $T_{th}$ on
element heterogeneity (i.e., $\sigma_I$) as well as the single
element NDR behavior. Interestingly, the pre-factor of the
$\ln\sigma_I$ term, calculated to be $\frac{C\,V_{max}}{\Delta
I}\simeq1.18\times10^{-4}$ s, has the form of an $RC$ time with an
effective resistance that depends only on the total current drop across
the NDR region, $\Delta I$. In particular, this time scale is not sensitive to the shape details of the $I_0(v)$ curve in the NDR region.  Figure 3c plots $T_{th}$ versus the
variance level (with the specific choice $r_{th}=0.1$) and shows
good agreement with numerical data for $N= 40$.

We have introduced a simple model that reveals how negative differential resistance and element
heterogeneity are key sources for the observation of non-uniform
field distributions and multiple current branches in the electrical
conduction properties of a series array of nonlinear circuit
elements. Specifically, we have shown how the system approaches a
state in which the element current levels are fully coupled as the element variance level and
voltage ramp rate are varied. The numerical and analytical results obtained
from this model provide insight for understanding similar
observed behaviors that are found for a range of more complex
electronic systems, for example, semiconductor superlattices and
quantum cascade laser structures. \cite{SLReview, LuS06,
Schoell2001} In practical devices it is often desirable to have an electric field distribution that is as spatially uniform as possible. The
model introduced here points out the relevance of two factors for achieving such field distributions.  One of these factors is the effective level of element heterogeneity in a periodic structure, which, in a device such as a superlattice, is determined in part by the fabricated sample quality and the electrical noise level.  The second factor suggests the possible use of a time-dependent voltage bias that possesses a ramping rate through the NDR region of the device that is large enough to maintain a nearly uniform field distribution.

We gratefully acknowledge helpful conversations with Luis Bonilla,
Gleb Finkelstein, Eckehard Sch\"{o}ll, and Adrienne Stiff-Roberts.  This work was supported by the National Science Foundation through Grant No. DMR-0804232.

\newpage

\begin{figure}[!h]
\centering
\includegraphics[width=14cm]{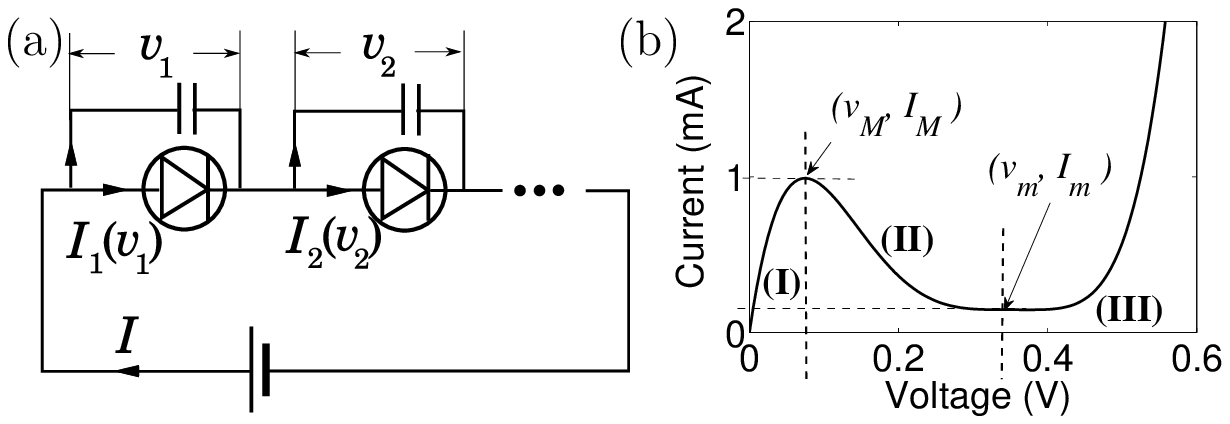}
\caption{(a) Nonlinear element array model.  Negative differential
resistance elements are connected in series with parallel
capacitance associated to each element. (b) The $I-V$ curve for a
typical element. Region (II) is the NDR region.}
\end{figure}

\begin{figure}[!h]
\includegraphics[width=14cm]{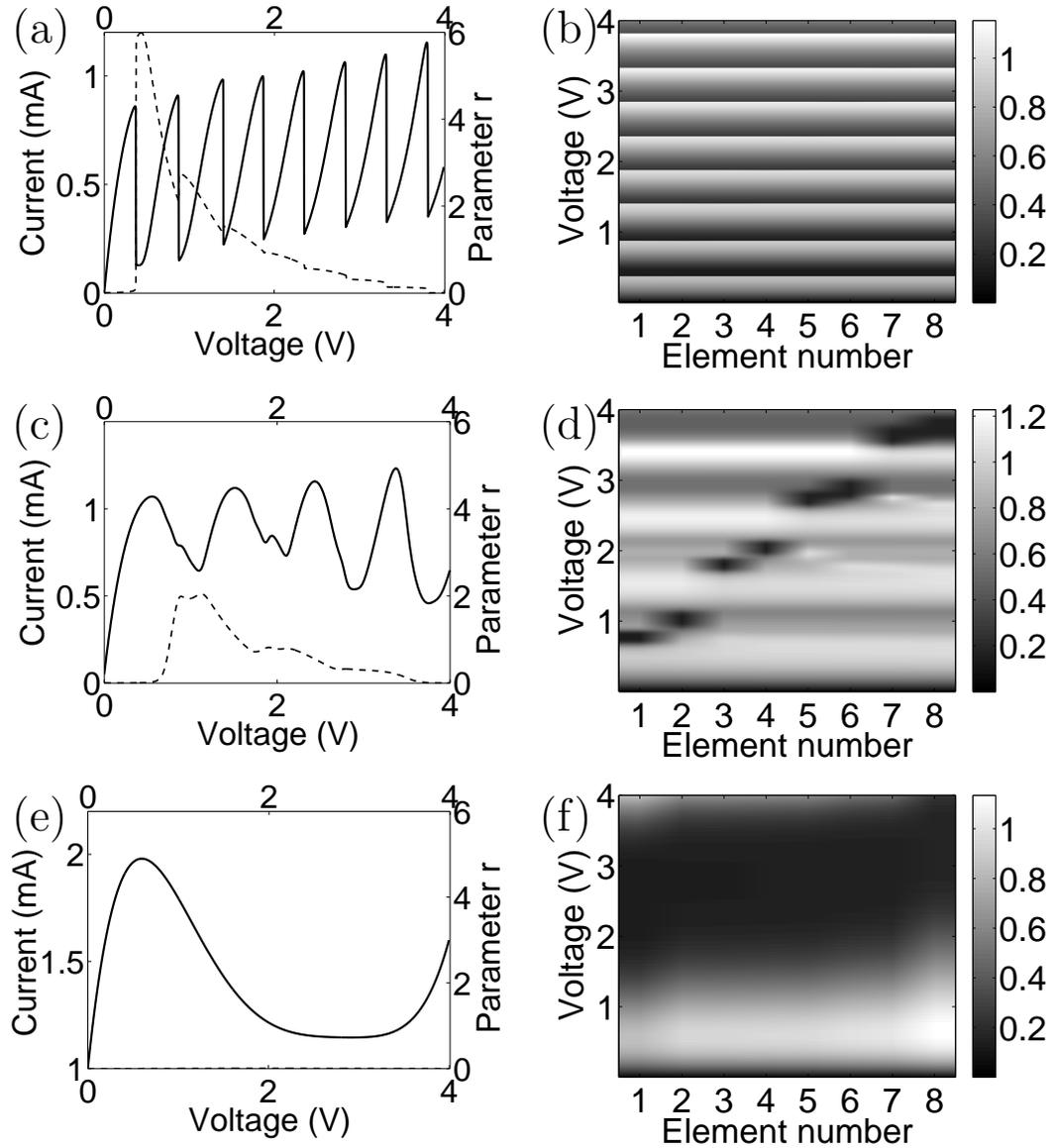}
\caption{Current-voltage ($I-V$) curves and current level contour plots
for different ramp times, all with $N = 8$, $C=200$ nF and
$\sigma_I=0.1$. (a) Fully coupled case: $I-V$ curve with ramp time
$T=500$ ms. Dashed curve shows the parameter $r(t)$. (b) Current contour plot for $T=500$ ms (gray scale in units mA). (c) $I-V$
curve for the partially coupled case with $T=2$ ms. (d) Current
contour plot for $T=2$ ms. (e) $I-V$ curve for the uncoupled case
with $T=0.1$ ms. (f) Current contour plot for $T=0.1$ ms.}
\end{figure}

\begin{figure}[!h]
\includegraphics[width=14cm]{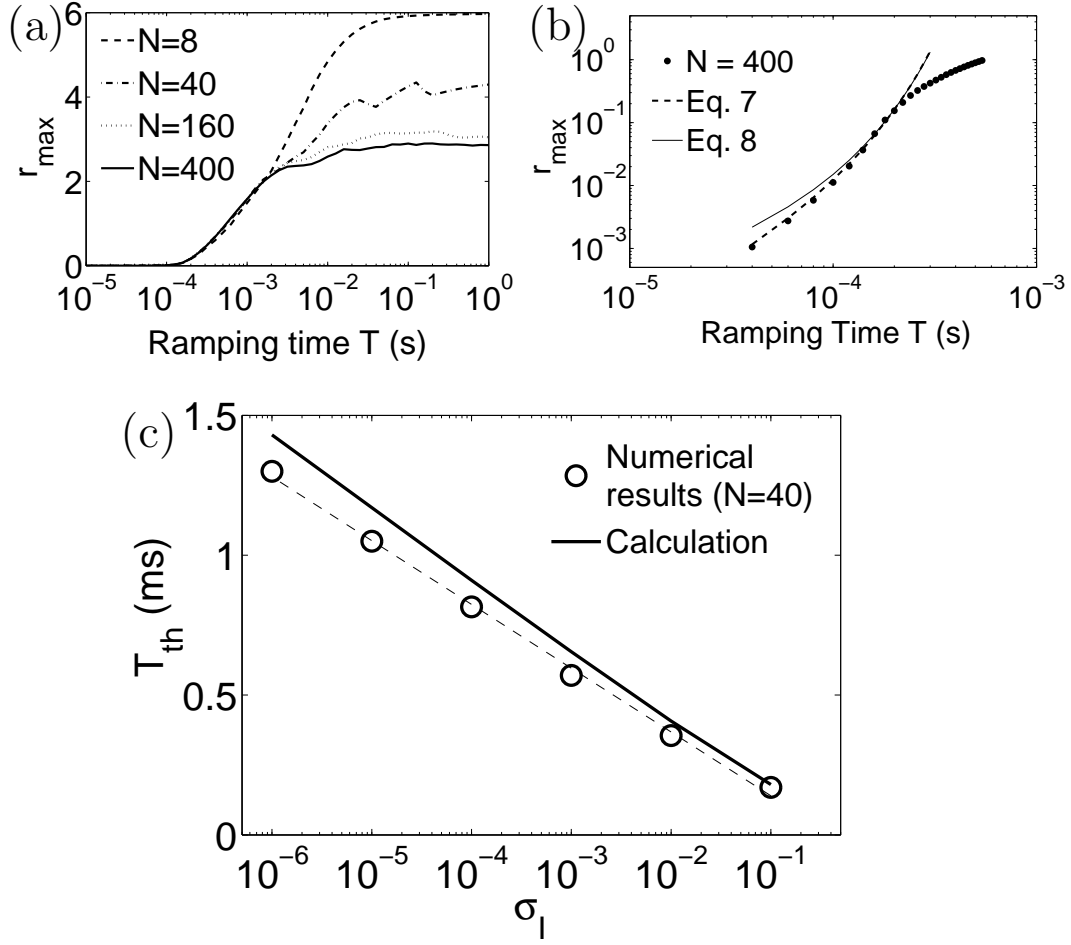}
\caption{(a) The parameter $r_{max}$ versus ramp time $T$ for
indicated values of $N$. Each curve is an average of ten different
statistical configurations of the $I_i(v_i)$ curves with $C=200$ nF
and $\sigma_I=0.1$. (b) Closed circles give $r_{max}$ vs. $T$ near
the \emph{characteristic} value for $N=400$. The solid curve is a
first-order calculation using Eqs.~(\ref{r}) and (\ref{dvcal}),
while the dashed curve is based on Eq.~(\ref{rmax}). (c) $T_{th}$
versus $\sigma_I$, comparing analytical prediction with simulation
for $N=40$.}
\end{figure}


\begin{thebibliography}{10}
\bibitem{footnote1}
For specificity, we use for the function $I_0(v)$ a fifth-order
polynomial fit to the measured $I-V$ curve of a tunnel diode (model
1N3712).
\bibitem{footnote2}
We use a Gaussian distribution on [-1,1] and have found that the
observed behavior is not sensitive to the detailed form of
distribution.
\end{thebibliography}
\end{document}